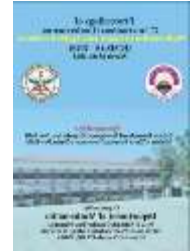

*RESEARCH ARTICLE*

# Modification in Elliptic Curve Cryptography based Mutual authentication scheme for smart grid communication using biometric approach

**Shivali[1*], Meenakshi[2]**
[1*,2]Srinivasa Ramanujan Department of Mathematics, Central University of Himachal Pradesh,
Dharmshala, 176215, Himachal Pradesh, India
*Corresponding Author E-mail: **koundalshivali7@gmail.com**

**ABSTRACT:**
Smart grid is a self-sufficient system. That tracks how the energy is used from its source to its final destination. The smart grid can increase the service quality while reducing the consumption of electricity. However, the safety and confidentiality of information/data is the major challenge in smart grid environment. To overcome this there are numerous authentication procedures that have been documented. The mutual authentication system for the smart grid that is based on elliptic curve cryptography and biometrics was thus introduced by A.A. Khan et al.'s. This protocol is secure from various attacks. But we found an inability of password and biometric updating phase. Therefore we provided the password and biometric updating phase in this protocol.

## 1. INTRODUCTION:

As the advancement of network technology and electrical technology smart grid plays an important role. In the past, smart grid used to deliver electricity to consumers homes and it is one-way transmission. But as the increasing power demands of 21$^{st}$ century it is difficult for one-way smart grid system to respond. Hence the two-way smart grid is introduced. In bidirectional smart grid system both electricity and information can interchange between the server and consumers. The smart grid system contains smart apparatus, grid sub-stations, phasor measurement and information transfers. Smart meters are used in smart grid system by smart equipments to exchange data with the server and the user. Additionally user inquiries are sent to substations via smart meters. Substations that received inquiries pass them to the control center, which then deals with the user's problems. In order to guarantee secure connection between users and substations while transmitting data safely and dependably, a number of authentication protocol have been established. However, these protocols have several limitations, such as the fact they cannot withstand certain attacks. Therefore, in a smart grid system authentication protocol is needed to secure this communication. Thus, cryptographic protocols are crucial to achieving a smart grid system's security and privacy. For smart grid system there are several schemes such as Biometric-based authentication scheme by Li and Hwang in 2010 [9]. In 2011 Mostafa M.F. et al. gave a simple message authentication mechanism for smart grid communication, which is based on the hash-based authentication code [10]. Further Chim et al. introduced a confidentiality-conserving authentication system for smart grid, in 2011. They use a privacy-preserving authentication technique employing pseudo identities and anti-counterfeit component at the slick instruments [11]. W.S. Juang et al. also gave a strong and effective smart card-based password-authenticated key agreement [13]. They makes significant advances by addressing the risk of smart card loss and by using elliptical curve methods to lower implementation cost and also states about the password updating phase. In 2009 D.-Z.Sun et al. introduced an improvement of password-authenticated key agreement using smart card [2]. In 2019, Km Renuka et al gave a decryption and development of a 3-factor authentication mechanism for wireless sensor systems that protects confidentiality. [4]. They examine the safety of a privacy-protection multi-factor authentication scheme for wireless sensor network. They also discuss the password and biometric updating phase. For application in a smart grid, H.P. Singh et al recommended the creation of a 3-factor user authentication system [12]. They revisit Wazid et al scheme and they states about the password updating phase. In 2021 D. Kaur et al. gave the decryption and development of a 2-factor authentication system for smart homes [3]. They also discuss the password updating phase. In 2021, A.A. Khan et al. gave a simple architecture for key establishment and authentication for the power grid [5]. They use random oracle model and also discuss the password and biometric updating phase in the protocol. Lastly, in 2017 Li et al. gave public key infrastructure based on network authentication mechanism for smart areas and buildings [8]. However, the expenses of processing are substantially higher.

### 1.1. Motivation:
To the best of our knowledge, the mutual authentication system for the smart grid that is based on ECC and biometrics requires a password and biometric updating phase. The biological and password update phase has been implemented into even the other authentication mechanisms. These factors inspire us to introduce the A.A. Khan protocol's biological and password update phase. The user can update or modify their biometric and password during this step.

### 1.2. Arrangement of a document:
The following is the order of a document: The fundamental representations utilized in this document are defined in part-2. After that, we review the A.A. Khan protocol's, in part-3. The part-4 is covered with the security evaluation of A.A. Khan's scheme. In section 5, we discuss our contribution and lastly we provide a conclusion.

### 2. Preliminaries:
This section provides the helpful notation and mathematical terms needed to understand the proposed scheme.

### 2.1. Fuzzy extractor and biometric:
The biometric information can be converted into a string of randomly generated characters with the help of fuzzy extractor and biometric are use to identifying the identity of user by using his/her fingerprints, face scan, voice recognition and iris scan.

### 2.2. Table:
**Notations and their definition.**

| Notations | Definition |
|---|---|
| ECC | Elliptic curve cryptography |
| $ID_i$ | person i's distinct identity |
| $PW_i$ | $i^{th}$ user's password |
| $B_i$ | $i^{th}$ user's biometrics |
| Gen(.) | Injector with fuzzy logic |
| S | Server |
| Rep(.) | Fuzzy manufacturing process |
| \|\| | Concatenation operator |
| + | XOR operation |
| → | Public channel |
| ⇒ | Secure channel |

- Gen:

A confidential data key $\sigma_i \in \{0,1\}^1$ and a common procreation variable $\tau_i$, are produced using a statistical technique that accepts a biometric input of $B_i \in N$, where $Gen(B_i) = \{\sigma_i, \tau_i\}$.

- Rep:

The parametric approach reproduces biometric key data $\sigma_i$ from noisy biometric data $B_I \in N$ and public parameter $\tau_i$, and T connected to $B_i$. Then $Rep(\dot{B}_i, \tau_i) = \sigma_i$.

### 3. Study of A.A. Khan Scheme:
The A.A. Khan Scheme is basically contains the three steps: initialization, enrollment, logon and authentication phase. Below is an explanation of these phases.

### 3.1. Startup stage:

The initial step: (server) S produce a hidden key X and shared key $PK_S = X.P$, where $P \in G$ and S choose p from $E_P(c,d): v^2 = u^3 + cu + d \bmod p$ with $c, d \in G$ and $4c^3 + 27a^2 \bmod p \neq 0$ and choose her/his $h(.)$. After that S reveal the public parameters $\{P, g, PK_S, h(.), E_P(c,d), p\}$ and secure the private key X.

### 3.2. Registration phase:

- In order to sign up with S, $U_i$ choose his $ID_i$, $PW_i$, imprint $B_i$, and evaluates $Gen(B_i) = (\sigma_i, \theta_i)$. Additionally, $U_i$ creates $r \in Z_q^*$, calculates $R_1 = h(PW_i \| \sigma_i) + r$ and sends $\{ID_i, R_i, t_{RG1}\}$ to the S via secure channel.
- The S verifies $t_{RG2} - t_{RG1} \leq \Delta t$ after getting $\{ID_i, R_i, t_{RG1}\}$. Then S evaluates $R_2 = h(ID_i\|X\|y)$ if the verification is successful. Where y is the counter. Additionally, the S evaluates $R_3 = R_2 + R_1$, stores $\{R_3, y, g, h(.)\}$ in the database and transmits it to $U_i$.
- $U_i$ evaluates $R_4 = R_3 + \sigma_i$, $R_5 = h(ID_i \| PW_i \| R_4)$ and stores $\{R_3, R_4, R_5\}$ in its database.

### 3.3. Login and authentication phase:

The user interacts with S in the domain of the smart grid at this phase.

- U enters the $ID_u'$, $PW_u'$, imprints $B_u'$. Additionally, U calculates, $\sigma_u' = Rep(B_u', \theta_U)$, $R_4' = R_3 + \sigma_u'$, $R_5' = h(ID_U\| PW_u\| R_4')$ and confirms $R_5' =^? R_5$. If so, U creates $u \in Z_q^*$, calculates $S_1 = h(ID_u \| R_1 \| t_1)$ and delivers $M_{A1} = \{S_1, ID_{u1}, u.g, t_1\}$ to S via secure channel.
- After receiving $M_{A1}$, S checks $t_2 - t_1 \leq \Delta t$, if successful, S evaluate $ID_u^* = ID_{u1} + (R_1 + t_1)$, $S_1^* = h(ID_u^*\| R_1 \| t_1)$ and then confirms that $S_1^* =^? S_1$. Then, S creates a random numbers $\in Z_q^*$, calculates $S_2 = h(ID_s \| R_3 \| t_2)$, $SK_{SU} = h(ID_u^*\| ID_S \| S_2 \| S_1^* \| R_3.g \| X.g \| u.s.g \| t_3)$ is the session key and determine the evaluates $ID_{S1} = ID_S + (R_3 + t_3)$. Additionally, S sends $M_{A2} = \{ID_{S1}, S_2, sg, t_3\}$ to U.
- U checks to see if $t_4 - t_3 \leq \Delta t$, if it is true, after receiving $M_{A2}$, U evaluates $ID_S^* = ID_{S1} + (R_3 + t_3)$, $S_2^* = h(ID_S^*\| R_3 \| t_3)$ and confirms that $S_2^* =^? S_2$. If so, make $SK_{US} = h((ID_u\| ID_S^* \| S_1 \| S_2^* \| R_3.g \| PK_S \|u.s.g \| t_3)$ as the session key.

### 4. Security analysis:

In this part, we go over the security evaluation of the aforementioned technique by A.A. Khan.

**4.1. Repeated assault:** S and U creates arbitrary values $u \in Z_q^*$ and $s \in Z_q^*$ and the time stamp circumstance $t_i - t_j \leq \Delta t$ at each phase. They ensure that the message will be current.

**4.2. Attacking man in the middle:** Hacker may attempt to log into the server using the prior messasge. Hacker replays $M_{A1} = \{S_1, ID_{u1}, u.g, t_1\}$, where $t_1$ is a time stamp, which stops a harm from replaying.

**4.3. User confidentiality:** In the aforementioned technique, U sends partial identity $ID_{U1}$ to S and S sends partial identity $ID_{S1}$ to U. Therefore attacker cannot get the original ID of U.

**4.4. Key freshness:** At each phase of the above scheme, uses a random number and a time stamp condition.

**4.5. Message authentication:** During the message authentication step, messages are secure within verification parameters and hash values that are difficult for an attacker to predict. The suggested approach therefore supports the message verification.

**4.6. False-identity assault:** Attacker might try to log in as an appropriate user U, receive message $M_{A1}$, which contains $S_1$, $ID_{U1}$, u.g and $t_1$ and attempt to compute $S_1$. However, this is difficult because $S_1$ consist of $ID_U$, $R_1$, $t_1$, and also $S_1$ is secure by biometric, random number and password.

**4.7. Key session deal:** According to the aforementioned technique, S and U compute their respective session keys as $SK_{SU}$ and $SK_{US}$. It is also evident that $SK_{SU} = SK_{US}$. Consequently, their communication is safe.

**4.8. Without trackability:** Every time the aforementioned approach is used, U chooses a new random number $u \in Z_q^*$ to calculate $M_{A1}$. There is no steady value sent by U because of its unpredictability. Hence, protocol offers non-trackability.

**4.9. Without mobility:** In the above scheme U check that $R_5^* =^? R_5$ during the login phase, this demonstrates that in the verification phase, only the authorized user can enter.

### 5. Our contribution:

The A.A. Khan et al.'s protocol is secure from various attacks as discuss above and have less communication cost, but we found that there is an inability of password and biometric updating phase. So we provided the biometric and password updating step in the A.A. Khan et al.'s protocol to make this protocol more efficient.

## 5.1. Biometric and password updating step:

The user can modify her/his biometric and password. Whenever, U needs to update her/his old password $PW_U$ and biometric $B_U$, just follows this instructions.

- User first enters $ID_U'$, $PW_U'$ and $B_U'$. Additionally, the U evaluates $\sigma_u' = Rep(B_u', \theta_U')$, $R_4' = R_3' + \sigma_u'$, $R_5' = h(ID_U \| PW_u \| R_4)$ and confirms $R_5' =^? R_5$; if this is false, the procedure is stopped. If not, the U chooses a new password $PW_U^*$ and biometric $B_U^*$, and then U evaluates $Gen(B_u^*) = (\sigma_u^*, \theta_u^*)$ and $R_1^* = h(PW_u^* \| \sigma_u^*) + r$ and sends $\{ID_U, R_1^*, t_{RG1}\}$ to the S via a secure channel.
- The S then evaluates $R_3^* = R_2^* + R_1^*$, stores $(R_3^*, y, g, h(.))$ in its database and delivers to U via a safe route.
- Then U evaluates $R_4^* = R_3^* + \sigma_u^*$ and $R_5^* = h(ID_U \| PW_u^* \| R_4^*)$.

Therefore, U changes $PW_U$ to $PW_U^*$, $B_U$ to $B_U^*$, $R_3$ to $R_3^*$, $R_4$ to $R_4^*$, $R_5$ to $R_5^*$, $\sigma_u$ to $\sigma_u^*$ and $\theta_u$ to $\theta_u^*$, and stores $\{R_3^*, R_4^*, R_5^*\}$ in its database.

**Table:**

| Password and biometric updating phase. | |
|---|---|
| User U | Server S |
| Enters $ID_U'$, $PW_U'$ and $B_U'$ | |
| Evaluates $\sigma_u' = Rep(B_u', \theta_U')$, $R_4' = R_3' + \sigma_u'$, | $R_5' = h(ID_U \| PW_u \| R_4)$ |
| Confirms $R_5' =^? R_5$; if yes: | |
| Selects $PW_U^*$, $B_U^*$ | |
| Evaluates $Gen(B_u^*) = (\sigma_u^*, \theta_u^*)$, | $R_1^* = h(PW_u^* \| \sigma_u^*) + r$ |
| Sends $\{ID_U, R_1^*, t_{RG1}\}$ | |
| ⟹ | evaluates $R_3^* = R_2^* + R_1^*$ |
|  | Store $(R_3^*, y, g, h(.))$ |
| ⟸ | ……...........sends |
| Evaluates $R_4^* = R_3^* + \sigma_u^*$ | |
| And $R_5^* = h(ID_U \| PW_u^* \| R_4^*)$ | |
| Stores $\{R_3^*, R_4^*, R_5^*\}$ in the database. | |

## 6. CONCLUSION:

As the advancement of network and electrical technology, smart grid is a crucial function, which interchanged both electricity and information between consumers and the server. The smart grid can increase the service quality while reducing the consumption of electricity. However, the security and privacy of communication/data is the major challenge in smart grid environment. A.A. Khan's suggested a mutual authentication system based on elliptic curve cryptography utilized biometric for the smart grid area [1], to safe the communication between the consumers and server. This system is less expensive to communicate with and secure against numerous assaults. However, we discovered that there was an inability of password and biometric updating phase, so we added it to A.A. Khan et al.'s protocol to improve its effectiveness. With the help of the biometric and password updating phase, you can modify the old biometric and password and keep your data safe even if an attacker or someone else discovers an old or preserved password.

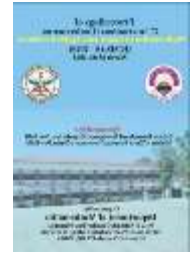

*RESEARCH ARTICLE*

# Cryptanalysis of LSPA-SGs: A lightweight and secure protocol for authentication and key agreement based Elliptic Curve Cryptography in smart grids

**Publish, Meenakshi**
Srinivasa Ramanujan Department of Mathematics, Central University of Himachal Pradesh,
Dharamshala (176215), India
*Corresponding Author E-mail: **thakurlovii0@gmail.com**

**ABSTRACT:**
Smart grids are becoming more and more significant as more nations adopt the smart city concept and boost energy sector efficiency to create a more sustainable and secure future. However, it is critical to address the security issues with smart grids. Security and privacy are essential components of SG communication. Recently, the LSPA-SGs scheme was created, and according to its creators, it is an effective and secure protocol. We reviewed their scheme and observed that it does not provide security and privacy. It contain some security vulnerabilities; user anonymity, stolen-verifier attack, password guessing attack, physical attack, privileged insider attack, user impersonation attack. This study exposed the weaknesses of Susan A. Mohammed et al design's and demonstrated how many security issues allowed for powerful attacks.

**KEYWORDS:** Smart grids, Elliptic curve cryptography, Authentication, key agreement, Security.

**1 INTRODUCTION:**
The first AC electric grid was built in Great Barrington, Massachusetts, in 1886 [1]. In this period, the distribution, transmission, and demand-driven regulation of energy were all handled by a single, consolidated grid. Local grids in the 20[th] century expanded throughout time and finally joined for practical and reliable reasons. Daily peaks in demand caused by residential heating and cooling were addressed by a variety of high-power generators that were only turned on briefly each day. Due to the low utilization of these peaking generators and the need for grid redundancy, gas turbines were typically used, which have lower capital costs and faster start up times. The electrical providers were hit with significant expenses as a result, which were subsequently passed on to customers in the form of higher prices. This electrical grid was not fulfilling the demands of 20[th] century populations due to a lack of natural gas, coal, water, and various fossil fuels, for which we had to introduce modern technology so that the electrical grid would become smarter. A better electrical power grid, a "smart grid," works with infrastructure communication technology to distribute electricity more effectively and to communicate with users and power communication providers. The 20th century's constantly evolving and expanding power needs cannot be met by the existing power grid architecture, making efficient power grid utilization essential today [1]. Among its many benefits, the smart grid allows for better management and expansion of renewable energy sources. Rapid advancements in communication and information technology in recent years have resulted in secure and ongoing technological advancements [2]. Just a few of the options it provides for developing a growing intelligent platform include power control, internet communication, and smart meters [3, 4]. A platform called SG enables two-way contact between users and service providers on a regular basis for computation and communication [5]. SG may be suspended from cyberattacks due to its sensitivity [6]. Physical attacks, cyberattacks, and natural disasters pose the